\documentclass[aps,prd,groupedaddress,showpacs,twocolumn]{revtex4}
\usepackage{amsmath}
\usepackage[parfill]{parskip}    % Activate to begin paragraphs with an empty line rather than an indent
\usepackage{float}
\usepackage{caption}
\usepackage{subcaption}
\usepackage{graphicx}
\usepackage{bbm}
\usepackage{graphicx}
\usepackage{amssymb}
\usepackage{epstopdf}
\usepackage{color}
\usepackage{ulem}
\usepackage{graphicx}
\usepackage{pdfpages}
\usepackage[colorlinks=true,citecolor=blue,urlcolor=magenta,breaklinks]{hyperref}
\begin{document}

\title{Light rings in static and extremal black holes}

\author{Pedro \surname{Bargueño}}
\email{pedro.bargueno@ua.es}
\affiliation{Departamento de F\'{i}sica Aplicada, Universidad de Alicante, Campus de San Vicente del Raspeig, E-03690 Alicante, Spain}

%%%%%%%%%%%%%%%%%%%%%%%%%%%%%%%%%%
\begin{abstract}
In this work we establish some results concerning the existence of external light rings in extremal black hole spacetimes through the Newman-Penrose formalism. Specifically, assuming conformal flatness, staticity and the null energy condition, we show that a sufficient condition for the existence of external light rings is $R<2 K_{G}$, where $R$, the curvature scalar of the spacetime and $K_{G}$, the Gaussian curvature of a spacelike two-surface, are
both evaluated at the outermost event horizon, which can be endowed with spherical, hyperbolic or planar geometry.
Our results are valid for any metric gravity theory where photons follow null geodesics.

\end{abstract}
%%%%%%%%%%%%%%%%%%%%%%%%%%%%%%%%%%%%%%%%%%

\maketitle

%%%%%%%%%%%%%%%%%%%%%%%%%%%%%%%%%%%%

\section{Introduction}
The ringdown and shadow observables are of fundamental importance to provide information on black hole geometry, which has been recently revealed
by gravitational wave observations \cite{Abbott1,Abbott2} and shadow images \cite{EHT1,EHT2,EHT3}. They are both intimately connected to a special set of bound null orbits for test particles \cite{Cardoso2016,Cunha2018} which, when planar, are known as light rings, an extreme form of light deflection consisting of closed paths.
\\
\\
The existence of null circular geodesics in generic (non-extremal) asymptotically flat black holes was proved in Ref. \cite{Hod2013} for spherically symmetric hairy configurations and for stationary axisymmetric black hole spacetimes \cite{Cunha2020}. The findings presented in Ref. \cite{Cunha2020} were extended to static and spherically symmetric black holes not only with asymptotically flat behavior, but also with (A)dS asymptotics \cite{Wei2020} and later to a general static warped product spacetime \cite{Koga2021}. A subsequent extension for light rings in a stationary spacetime with an ergoregion was presented in \cite{Ghosh2021}. These recent works, mainly based on topological and/or effective potential techniques, were
recently followed by a more geometric approach for spherically symmetric, static and non-extremal spacetimes, based on the Gauss and geodesic curvatures of the optical metric \cite{Qiao2022a,Qiao2022b,Cunha2022}. Interestingly, although some universal properties of light rings for stationary and axisymmetric spacetimes, including the extension to the extremal case, were recently reported \cite{Guo2021}, the static case remains essentially open 
\footnote{After the completion of this work, the existence of external null circular geodesics in spherically symmetric black holes has been reported by Y. Peng in arXiv:2211.14463. In addition, S. Hod has shown in arXiv:2211.15983 that
spherically symmetric extermal black holes possess at least one external light when the dominant energy condition and Einstein's field equations are assumed.}. In this sense, we would like to mention that the problem of the existence of light rings in static, spherically symmetric and asymptotically flat extremal black holes was considered in Ref. \cite{Hod2022} within the framework of general Relativity, obtaining that it crucially depends on the sign of the tangential pressure of the matter sector. 
\\
In the present work we shall focus our attention on light rings for general static and extremal black holes with different horizon topologies and Minkowskian or (A)dS asymptotics, irrespective of the underlying gravitational theory. Therefore, our results will be valid for theories beyond General Relativity. The manuscript is organized as follows: section \ref{sect2} is devoted to prove 
our main results using the Newman-Penrose formalism, whose essentials are introduced at this point. Discussions and final remarks are left to section \ref{sect3}.

\section{Light rings through the Newman-Penrose calculus}
\label{sect2}
Here we will follow Penrose and Rindler's conventions \cite{Penrose1984}. Let us start from a spherically symmetric and static geometry, written as $ds^2= f(r)dt^2-g(r) dr^2-r^2 d\Omega^2$, where $d\Omega^2=d\theta^2 +\sin^{2} \theta d\phi^2$. After choosing the following null tetrad
\begin{eqnarray}
l^{\mu}&=&\left(\frac{1}{\sqrt{2 f}},-\frac{1}{\sqrt{2 g}},0,0\right), \nonumber \\
n^{\mu}&=&\left(\frac{1}{\sqrt{2 f}},\frac{1}{\sqrt{2 g}},0,0\right), \nonumber \\
m^{\mu}&=&\left(0,0,-\frac{1}{\sqrt{2}r},\frac{\textrm{i}\csc \theta}{ \sqrt{2}r}\right),
\end{eqnarray}

where $l_{\mu}n^{\mu}$=1 and $m_{\mu}\bar m^{\mu}=-1$, with the bar denoting complex conjugation, the only non--vanishing Newman-Penrose scalars for the considered spacetime (which will be either Petrov type D or O due to spherical symmetry) are 
\\
\begin{eqnarray}
\Psi_{2}&=&C_{pqrs}l^{p} m^{q} \bar m^{r} n^{s} \nonumber \\
\Phi_{11}&=&-\frac{1}{2}R_{ab}l^{a}n^{b}+3 \Lambda \nonumber \\
 \Phi_{00}&=& -\frac{1}{2}R_{ab}l^{a}l^{b} \nonumber \\
 \Phi_{22}&=& -\frac{1}{2}R_{ab}n^{a}n^{b} \nonumber \\
\Lambda&=&\frac{R}{24}, \nonumber \\
\end{eqnarray}
\\
where $C_{abcd}$ and $R_{ab}$ stand for the Weyl and Ricci curvatures, respectively, and $R= g^{a b}R_{a b}$.arXiv:2211.15983 (2022).

In particular, for the geometry under consideration, we obtain
\begin{widetext}
\begin{eqnarray}
\label{NP}
\Psi_{2}&=& \frac{f''}{12 f g}-\frac{f' g'}{24 f g^2}-\frac{f'^2}{24 f^2 g}-\frac{f'}{12 r f g}+\frac{g'}{12 r g^2}+\frac{1}{6 r^2 g}-\frac{1}{6 r^2}
\nonumber \\
\Phi_{11}&=&\frac{f''}{8 f g}-\frac{f' g'}{16 f g^2}-\frac{f'^2}{16 f^2 g}-\frac{1}{4 r^2 g}+\frac{1}{4 r^2}
\nonumber \\
R&=&-\frac{f''}{f g}+\frac{f' g'}{2 f g^2}+\frac{f'^2}{2 f^2 g}-\frac{2 f'}{r f g}+\frac{2 g'}{r g^2}-\frac{2}{r^2 g}+\frac{2}{r^2} \nonumber \\
\Phi_{00}&=&\frac{\Phi_{22}}{16}=\frac{4 \left(g f'+f g'\right)}{r f g^2},
\end{eqnarray}
\end{widetext}
where the prime denotes derivative w. r. t. the radial coordinate.
Interestingly, in the case $\Phi_{00}=\Phi_{22}=0$, which implies $f(r) g(r) = A$ (we always can take $A=1$ by a re-parametrization of the time coordinate) we can solve for the metric potential and its derivatives in terms of the Newman Penrose scalars, obtaining
\\
\\
\begin{eqnarray}
\label{fs}
f&=&1-2r^2\left(\Lambda+\Phi_{11}-\Psi_{2}\right) \nonumber \\
f'&=&-2r\left(2\Lambda+\Psi_{2}\right)\nonumber \\
f''&=&4\left(-\Lambda+\Phi_{11}+\Psi_{2}\right).
\end{eqnarray}
\\
\\
At this point, let us introduce the light ring condition as follows: a black hole has an external light ring located at $r_{\gamma}$, where $r_{\gamma}>r_{h}$ being $r_{h}$  the location of the outermost event horizon, when 
\\
\begin{equation}
\label{Dfunction}
    D(r_{\gamma})= r_{\gamma} f'(r_{\gamma})- 2 f(r_{\gamma})=0.
\end{equation}
\\
For our purposes, we rewrite the $D$-function appearing in Eq. (\ref{Dfunction}) as
\\
\begin{equation}
    D(r)=-2+2r^2(2 \Phi_{11}-3 \Psi_{2}),
\end{equation}
or
\begin{eqnarray}
    \frac{D}{2r^2}&=&-K_{G}+2 \Phi_{11}-3 \Psi_{2} \nonumber \\
    \label{Dbis}
    &=&-K_{G}+2 \left(\Phi_{11}+3 \Lambda \right)-3 \left(\Psi_{2}+2\Lambda \right),
\end{eqnarray}
where $K_{G}=\frac{1}{r^2}$ stands for the Gaussian curvature of the angular sector of the spacetime (which is a 2-sphere in the case here
considered).
\\
\\
Let us remark that the function $D(r)$ goes to -2 not only for an asymptotically flat spacetime, but also for (A)dS asymptotics (which are conformally flat), where $\Psi_{2}=0$. 
\\
\\
The first of Eqs. (\ref{fs}), which we refer to the Penrose-Rindler equation, will be useful along the manuscript. It can be expressed as
\\
\begin{eqnarray}
    \frac{f}{2r^2}&=&\frac{K_G}{2}-\Lambda-\Phi_{11}+\Psi_{2} \nonumber\\
    &=& \frac{K_G}{2}-\left(\Phi_{11}+3 \Lambda\left) +\right(\Psi_{2}+2 \Lambda\right). 
\end{eqnarray}
At this point, a couple of comments are in order:
\begin{itemize}
    \item The dominant (null) energy condition, together with Einstein's equations, imply $\Phi_{11}+3\Lambda \ge 0$ ($\Phi_{00}\ge 0$) \cite{Witt1973}.
    \item $ \frac{K_G}{2}-\left(\Phi_{11}+3 \Lambda\right) +\left(\Psi_{2}+2 \Lambda\right)=0$ for both extremal or non-extremal black hole event horizons.
    \item An extremal black hole is characterized by $f'(r_{h})=0$ which, in the $\Phi_{00}=0$ case, is equivalent to $\left(\Psi_{2}+2\Lambda \right)|_{r_{h}} =0$.
    \item $\frac{K_G}{2}-\left(\Phi_{11}+3 \Lambda\right)=0$ for an extremal black hole horizon.
    \end{itemize}

Therefore, we can conclude that $D=-3 \left(\Psi_{2}+2\Lambda \right)>0$ for a non-extremal black hole. Then, having into account the $D\rightarrow -2$ asymptotic limit,  at least one external light ring is shown to exist. This results generalizes previous findings \cite{Cunha2022} by including (A)dS asymptotics.
\\
\\
Note that the condition $D'(r_{h})> 0$ implies the existence of at least one external light ring for extremal black holes, which satisfy $D(r_{h})=0$.
In fact, a straightforward calculation reveals that, for an extremal black hole horizon,
\\
\begin{eqnarray}
    D'(r_{h}) &=& 2 r_{h} \left( 3 \Psi_{2} + 2 \Phi_{11} \right)|_{r_{h}} \nonumber \\
    &=& 4 r_{h} \left(\Phi_{11} -3 \Lambda\right)|_{r_{h}}  \nonumber \\
    &=& 4 r_{h} \left(2 K_{G}-R\right)|_{r_{h}} .
\end{eqnarray}
Therefore, we obtain the following
\\
\\
{\it Theorem 1.} Let us consider an asymptotically conformal
based on the dominant energy condition which characterizes the external matter fields in non-vacuum extremal black hole spacetimes, .  
\\ly flat, spherically symmetric, static and extremal black hole spacetime with $\Phi_{00}=0$. If $R(r_{h})<2 K_{G}(r_{h})$, then there is at least one external light ring. 
\\
\\
In particular, the existence of external light rings for extremal black holes with $R(r_{h})\le 0$ is guaranteed.
\\
\\
Our results can be extended to the $\Phi_{00}\ne0$ case. Recall that $f'(r_{h})=0$ also for an extremal black hole. In this case, we get from Eqs. (\ref{NP})
that, at the horizon of an extremal black hole:
\begin{eqnarray}
    \left(\Phi_{11}+3\Lambda\right) &|_{r_{h}} =& \frac{1}{2 r_{h}^2}+\frac{g'(r_{h})}{4 r_{h} g(r_{h})^2} \nonumber\\
    \Phi_{00}|_{r_{h}} &=& \frac{ 4 g'(r_{h})}{r_{h} g(r_{h})^2} \nonumber \\
    R|_{r_{h}}  &=& \frac{2}{r_{h}^2}+\frac{2 g'(r_{h})}{r_{h} g(r_{h})^2} -\frac{f''(r_{h})}{f(r_{h}) g(r_{h})},
\end{eqnarray}
\\
and, therefore, 
\begin{equation}
    D'(r_{h})=r_{h} f''(r_{h}) = \left(2 K_{G}+\frac{\Phi_{00}}{2}-R\right)|_{r_{h}} .
\end{equation}
At this point, we take take advantage of a result by Hayward concerning {\it trapping horizons} \cite{Hayward1994a,Hayward1994b} which, in 
the spherically symmetric case, can  be re-stated as follows:
\\
\\
{\it Signature law}. If the null energy condition holds on a spherically symmetric trapping horizon, $r_{t}$, the horizon is null if and only if 
$\Phi_{00}(r_{t})=0$.
\\
\\
But, in the spherically symmetric and static case, both event and trapping horizons coincide \cite{Nielsen2006,Nielsen2010}, $r_{t}=r_{h}$. 
Therefore, as any event horizon is null, the signature law implies that, if the null energy condition holds on $r_{h}$, then $\Phi_{00}(r_{h})=0$. Therefore, we can state the following
\\
\\
{\it Theorem 2.} Let us consider an asymptotically conformally flat, spherically symmetric, static and extremal black hole spacetime. If the null
energy condition holds on the outermost event horizon, $r_{h}$, and $R(r_{h})<2 K_{G}(r_{h})$, then there is at least one external light ring. In particular, the existence of this external light ring is guaranteed if $R(r_{h})\le 0$.
\\
\\
Finally, we would like to point out that our results can be easily extended to the case of hyperbolic and planar horizons. 
The line element is written as $ds^2= f(r)dt^2-g(r) dr^2-r^2\left(d\theta^2+\gamma^2(\theta)d\phi^2\right)$, where $\gamma(\theta)=\{1,\sinh \theta, \sin \theta\}$ for planar, hyperbolic and spherical horizons, respectively. For these metrics, both $l^{\mu}$ and $n^{\mu}$ are independent of the geometry
of the angular sector, but $m^{\mu}=\left(0,0,-\frac{1}{\sqrt{2}r},\frac{\textrm{i}}{ \gamma(\theta)\sqrt{2}r}\right)$. 
\\
\\
Within this general situation, a straightforward calculation reveals that Eqs. (\ref{fs}) can be expressed as
\\
\begin{eqnarray}
\label{fsbis}
f&=&\alpha-2r^2\left(\Lambda+\Phi_{11}-\beta \Psi_{2}\right) \nonumber \\
f'&=&-2r\left(2\Lambda+\beta \Psi_{2}\right)\nonumber \\
f''&=&4\left(-\Lambda+\Phi_{11}+\beta \Psi_{2}\right),
\end{eqnarray}
\\
where $(\alpha,\beta)$ is $(0,1)$, $(-1,1)$ and $(1,1)$ for planar, hyperbolic and spherical horizons, respectively. Therefore, following the same
arguments we can state the following
\\
\\
{\it Theorem 3.} Let us consider an asymptotically conformally flat, static and extremal black hole spacetime. If the null
energy condition holds on the outermost event horizon, $r_{h}$, and $R(r_{h})<2 K_{G}(r_{h})=\frac{2 \alpha}{r_{h}^2}$, then there is at least one external light ring. \\

\section{Discussion and final remarks}
\label{sect3}
At this point, the results we have obtained are valid for any metric theory of gravity where photons follow null geodesics. In the particular case of General Relativity, including a non-vanishing cosmological constant, $\lambda$, the aforementioned sufficient condition reads
\begin{equation}
\label{trace}
    T < \frac{K_{G}}{4 \pi}-\frac{\lambda}{2\pi},
\end{equation}
where $T$ stands for the trace of the energy-momentum tensor. Note that the trace energy condition, the assertion that the
trace of the stress-energy tensor should (in mostly minus signature) be non-negative it has now been completely
abandoned and is no longer cited in the literature \cite{Barcelo2002}. In this sense, there are not restrictions of this kind with respect to Eq. (\ref{trace}). 
\\
\\
Finally, we can easily recover a very recent result \cite{Hod2022} concerning light rings in extremal, static and spherically symmetric black holes within the framework of General Relativity under the dominant energy condition. Specifically, if the dominant energy condition holds, then $\Phi_{00}(r_{h})=0$. Then, if the Einstein equations are assumed, $2K_{G}(r_{h})-R(r_{h}) >0$ implies $-8 \pi \Big(\rho(r_{h})-p_{r}(r_{h})-2 p_{t}(r_{h}) \Big)<\frac{2}{r_{h}^2}$. Note that $\Phi_{00}(r_{h})=0$ implies, by Einstein's equations, $\rho(r_{h})+p_{r}(r_{h})=0$ and, therefore, we get $p_{r}(r_{h})+p_{t}(r_{h})>\frac{1}{8 \pi r_{H}^2}$. But, as shown in Ref. \cite{Mayo1996}, $1-  8 \pi r_{h}^2 p_{r}(r_{h})=0$ for an extremal black hole. Then, $R(r_{h})<2 K_{G}(r_{h})$ implies $p_{t}(r_{h})<0$, which coincides with the main result of Ref. \cite{Hod2022} when our mostly minus signature is switched to the mostly plus choice of \cite{Hod2022}.
\\
\\
Summarizing, we have successfully employed the Newman-Penrose formalism to tackle the problem of the existence of external light rings in static an extremal black holes with planar, hyperbolic and spherical horizons, deriving a sufficient condition expressed in terms of the scalar curvature of the whole spacetime and the Gaussian curvature of the horizon. We have extended previous results in spherical symmetry by including conformally
flat asymptotics and, although our results do not depend on the underlying metric gravitational theory, we have recovered very recent results concerning extremal black holes within General Relativity using the Newman-Penrose formalism.

\section{Acknowledgements}

P. B. is funded by the Beatriz Galindo contract BEAGAL 18/00207, Spain. P. B. acknowledges Ana\'{\i}s, Luc\'{\i}a, In\'es and Ana for continuous support.

\end{document}